\documentclass[dvipdfmx,12pt]{article}
\usepackage[dvipsnames]{xcolor}
\usepackage{amsmath} 
\usepackage{amssymb}
\usepackage{mathtools}
\usepackage{physics}
\usepackage{bm}
\usepackage[margin =27truemm]{geometry}
 \usepackage{cite}
 
\begin{document}
\setlength{\baselineskip}{0.6cm}

\begin{titlepage}
\begin{flushright}
NITEP 242
\end{flushright}

\vspace*{10mm}%

\begin{center}{\Large\bf
Family Unification in a Six Dimensional Theory \\
\vspace*{2mm}
with an Orthogonal Gauge Group
}
\end{center}
\vspace*{10mm}
\begin{center}
{\large Nobuhito Maru}$^{a,b}$ and
{\large Ryujiro Nago}$^{a}$ 
\end{center}
\vspace*{0.2cm}
\begin{center}
${}^{a}${\it
Department of Physics, Osaka Metropolitan University, \\
Osaka 558-8585, Japan}
\\
${}^{b}${\it Nambu Yoichiro Institute of Theoretical and Experimental Physics (NITEP), \\
Osaka Metropolitan University,
Osaka 558-8585, Japan}
\end{center}

\vspace*{20mm}


\begin{abstract}
We propose a simple model of family unification, which is a six dimensional $SO(20)$ gauge theory 
with a single fermion in the spinorial representation. 
After compactification to five dimensions, our model gives a five dimensional model 
where the Standard Model Higgs field is unified into the fifth component of the five dimensional gauge field 
as well as three generations of quarks and leptons are unified into a single spinor field. 
\end{abstract}

\end{titlepage}

The origin of three generations of quarks and leptons is one of the mysteries in the Standard Model (SM). 
An idea to solve this problem is to extend the SM gauge group or the grand unified gauge group 
to a larger symmetry group $G$ incorporating the structure of three generations. 
The three generations of SM fermions are embedded into ideally a single fermion 
or non-repetitive set of fermions in some irreducible representations of $G$. 
After symmetry breaking of the group $G$ to the SM group or the grand unified group, 
three generation of the SM fermions are obtained as massless fields. 
This approach is often called as ``Family unification". 
There have been many proposals for the family unification from the various viewpoints, so far  
\cite{
Georgi, Frampton1, FN, Frampton2, IKK, WZ, NSS, Fujimoto, BDM, BSMR, Barr, FK1, BBK, 
HS, KKO, KK, FK2, KM, AFK, GKM, GK, RVVW, Yamatsu, CLY
}. 

In our previous paper \cite{MN}, we have proposed a six dimensional (6D) $SU(14)$ model of family unification 
based on the grand gauge-Higgs unification (GGHU) where three generation quarks and leptons are obtained 
from several fermions in totally anti-symmetric representations. 
A remarkable point in this model is that the SM gauge fields and Higgs field are unified in the higher dimensional gauge field  
as well as the the unification of three generations of the SM fermions. 
However, there also remained some unsatisfactory points. 
First, several fermions were necessary for obtaining three generations. 
As mentioned above, it is desirable to obtain three generations from a single fermion. 
Second, some unwanted massless fermions not the SM fermions were often left after symmetry breaking. 
To provide masses for them, we have to introduce some additional interactions and fields, 
then the model becomes complicated. 

In this letter, we propose a new simple model of family unification avoiding the above unsatisfactory points. 
This model is also based on the GGHU, but an orthogonal group is employed as a gauge group. 
As was discussed in \cite{MN}, the reason of considering the orthogonal group is that 
a spinor representation in the orthogonal group has a very large dimension 
with respect to the size of the gauge group. 
In fact, three generation of the SM fermions can be obtained from a single fermion belonging to the spinorial representation 
by considering some simple dynamical assumptions. 
We emphasize here again that the SM Higgs field is also unified in the higher dimensional gauge field, 
which was not considered in 
\cite{
Georgi, Frampton1, FN, Frampton2, IKK, WZ, NSS, Fujimoto, BDM, BSMR, Barr, FK1, BBK, 
HS, KKO, KK, FK2, KM, AFK, GKM, GK, RVVW, Yamatsu, CLY
}. 
Our strategy of the model building is as follows. 
We start from a 6D SO gauge theory with a single fermion in a spinorial representation. 
By compactifying the sixth spatial dimension on an orbifold $S^1/Z_2$, 
we obtain a model of 5D gauge-Higgs unification incorporating three generation structure. 

Let us consider a 6D $SO(20)$ gauge theory with a single fermion in a spinorial representation ${\bf 1024}$, 
which is compactified on an orbifold $S^1/Z_2$ in a sixth dimension with a radius $R_6$. 
In this model, we have no perturbative and nonperturbative gauge anomalies 
since 6D spinor in ${\bf 1024}$ is Dirac for the perturbative anomaly 
and the sixth homotopy group of $SO(20)$ is zero $\Pi_6(SO(20))=0$ for the nonperturbative anomaly.  
 
First, the $Z_2$ parity transformations for the gauge field are given as follows.   
\begin{align}
A_M(x_\mu, x_5, x_i-x_6) &= P_i A_M(x_\mu, x_5, x_i + x_6) P_i^\dag~(M=0,1,2,3,5), \\
A_6(x_\mu, x_5, x_i-x_6) &= - P_i A_6(x_\mu, x_5, x_i + x_6) P_i^\dag,
\label{Z26Dgauge}
\end{align}
where $P_i~(i=0,1)$ are parity matrices around the fixed points $x_0=0, x_1=\pi R_6$, 
which are chosen so that the desirable gauge symmetry may be remained.

$SO(20)$ generators are represented as follows.
\begin{align}
\sigma_2 \otimes S_{10}, \quad \sigma_0 \otimes A_{10}, \quad \sigma_2 \otimes I_{10},  \quad
\sigma_1 \otimes A_{10}, \quad \sigma_3 \otimes A_{10},
\label{SO(20)}
\end{align}
where $\sigma_i (i=1,2,3)$ are Pauli matrices, 
$S_N (A_N)$ is $N \times N$ (anti-)symmetric matrices, respectively.  
If we choose a $Z_2$ parity matrix at $x_6=0$ as 
\begin{align}
P_0 =\sigma_0 \otimes I_{6,4},
\end{align}
where $\sigma_0 = \rm{diag}(1, 1)$ and $I_{6,4} \equiv \rm{diag}(1,1,1,1,1,1,-1,-1,-1,-1)$, 
then, the generators corresponding to the unbroken gauge symmetry commute with this $P_0$. 
The result is 
\begin{align}
&\sigma_2 \otimes S_6, \quad \sigma_0 \otimes A_6, \quad \sigma_2 \otimes I_6, 
\quad \sigma_2 \otimes I_4, \quad \sigma_2 \otimes S_4, 
\notag \\
&\sigma_0 \otimes A_4, \quad 
\sigma_1 \otimes A_6, \quad\sigma_1 \otimes A_4, \quad \sigma_3 \otimes A_6, \quad \sigma_3 \otimes A_4. 
\label{SO(12)SO(8)}
\end{align}
In this case, the gauge symmetry $SO(20)$ is broken to $SO(12) \times SO(8)$ \cite{WZ, KM}. 

If we further choose a $Z_2$ parity matrix at $x_6=\pi R_6$ as 
\begin{align}
P_1=( I_{11,1}, \sigma_2 \otimes I_4),
\end{align}
where $I_4$ is a $4 \times 4$ unit matirx, 
the unbroken gauge symmetry is $SO(11) \times SU(4) \times U(1)$ and 
$U(1)$ denotes a subgroup of $SO(8)$, respectively. 
This is because 
the generators commutable with $\sigma_2 \otimes I_4$ are 
$SU(4)$ generators $\sigma_2 \otimes S_4, \sigma_0 \otimes A_4$ and a $U(1)$ generator $\sigma_2 \otimes I_4$ 
in $SO(8)$ generators 
$\sigma_2 \otimes S_4, \sigma_0 \otimes A_4, 
\sigma_2 \otimes I_4, \sigma_1 \otimes A_4, \sigma_3 \otimes A_4$. 

A 6D Dirac $SO(20)$ spinorial representation ${\bf 1024}$ can be decomposed 
as ${\bf 1024} = {\bf 512}_+ + {\bf 512}_-$, 
where $\pm$ means $Z_2$ parity eigenvalues of $\Gamma^{21}$ in $SO(20)$.  
We introduce only ${\bf 512}_+$ to our model. 
We note that $512_+$ is 6D Dirac fermion, which has eight components.
$Z_2$ parity transformations $x_i - x_6 \to x_i + x_6$ of 6D Dirac fermion are given
\begin{align}
{\bf 512}_+ (x_i - x_6) =  \eta_i P_i (i\Gamma^6) {\bf 512}_+ (x_i + x_6),  
\end{align}
where $\eta_i$ is a sign ambiguity $\pm 1$. 
Now, we take $(\eta_0, \eta_1) = (+, +)$ so that the four components of ${\bf 512}_+$ 
have $+1$ eigenvalues of $i \Gamma^6$ and remain to be massless. 
In the symmetry breaking $SO(20) \to SO(11) \times SU(4) \times U(1)$ by $P_0$ and $P_1$, 
${\bf 512}_+$
is decomposed into the following irreducible representations \cite{Group1, Group2}.
\begin{align}
\bm{512}_+ &= (\bm{32}, \bm{8}_s) \oplus (\bm{32}, \bm{8}_c) 
\nonumber \\
&= [(\bm{32}, \bm{1}_2) \oplus  (\bm{32}, \bm{1}_{-2}) \oplus (\bm{32}, \bm{6}_0)] + 
[(\bm{32}, \bm{4}_{-1}) \oplus  (\bm{32}, \bm{\overline{4}}_{1})],
\end{align}
where 
the representations in the parentheses are those of $SO(11) \times SO(8)$ in the first line, 
$SO(11) \times SU(4) \times U(1)$ in the second line and the number in the subscript implies a $U(1)$ charge. 
The subscripts $s, c$ of ${\bf 8}$ mean the spinor and conjugated spinor representations of $SO(8)$ 
\cite{Group1, Group2}\footnote{In general, the $2^{n-1}$ dimensional (conjugated) spinor of $SO(2n)$ can be decomposed 
into the sum of the (odd) even number rank totally anti-symmetric tensors on $SU(n)$, respevtively.   
In our case, the branching rules of 8 dimensional spinor and conjugated spinor representations 
under $SU(4) \times U(1)$ are
$\bm{8}_s = [0] + [2] + [4] = \bm{1}_2 + \bm{6}_0 + \bm{1}_{-2}, \bm{8}_c = [1] + [3] = \bm{4}_{-1} + \bm{\overline{4}}_{1}$, 
where $[n]$ represents the $n$-rank totally anti-symmetric tensor of $SU(4)$. 
See the page 204 of \cite{Group2}, for instance.}. 

If we assume that a subgroup $Sp(4)$ of $SU(4)$ is a confining gauge group as QCD in 
the SM \cite{WZ}\footnote{$SU(4) \to Sp(4)$ can be possible 
if we introduce a scalar field in the representation ({\bf 1}, {\bf 6}, any non-zero charge) 
under $SO(11) \times SU(4) \times U(1)$ unbroken at $x_6=\pi R_6$ fixed point 
and we construct the double-well potential of the scalar field such as the Higgs potential in the SM, which realizes its VEV. 
}~
the corresponding decomposition of spinors into the irreducible representations of $Sp(4)$ are as follows \cite{Group1, Group2}. 
\begin{align} 
&(\bm{32}, \bm{8}_s) = (\bm{32}, \bm{1}_2) \oplus  (\bm{32}, \bm{1}_{-2}) \oplus (\bm{32}, \bm{1}_0) 
\oplus (\bm{32}, \bm{5}_0), 
\label{Sp4_1} \\
&(\bm{32}, \bm{8}_c) = (\bm{32}, \bm{4}_{-1}) \oplus  (\bm{32}, \bm{4}_{1}). 
\label{Sp4_2}
\end{align}
Since the massless spinors should correspond to the $Sp(4)$ singlets, 
we find three massless $\bm{32}$ spinors, which will provide three generations of quarks and leptons 
in a model of 5D $SO(11)$ gauge-Higgs grand unification. 
In the compactification on the orbifold $S^1/Z_2$ to 4D, 
we can impose $Z_2$ parity such that the left-handed Weyl fermion ${\bf 16}_L$ 
is massless, 
where three generations of the SM fermions are only realized in the zero mode sector 
and no other massless fermions are present. 

As for the 6D massless gauge field $A_M~(M=0,1,2,3,5)$ of $SO(20)$ in ${\bf 190}$ dimensional adjoint representation, 
it is decomposed under $SO(12) \times SO(8)$ and $SO(11) \times SU(4) \times U(1)$ symmetry breaking as \cite{Group1, Group2}
\begin{align}
{\bf 190} &= ({\bf 66}, {\bf 1}) +({\bf 1}, {\bf 28}) + ({\bf 12}, {\bf 8}_v)~(SO(12) \times SO(8)), \\
&= ({\bf 55} + {\bf 11}, {\bf 1}_0) +({\bf 1}, {\bf 15}_0 + {\bf 6}_2 + {\bf 6}_{-2} + {\bf 1}_0), \nonumber \\ 
& \hspace*{30mm} + ({\bf 11} + {\bf 1}, {\bf 4}_{-1} + \overline{{\bf 4}}_{1})~(SO(11) \times SU(4) \times U(1))
\end{align}
${\bf 8}_v$ is a vector representation of $SO(8)$. 
\begin{table}[h]
    \centering
    \begin{tabular}{c|c}
   $(SO(11), SU(4))_{U(1)}$ & ($P_0, P_1$)\\
    \hline
    (${\bf 55}$, ${\bf 1}_0$) & (+, +) \\
    (${\bf 11}$, ${\bf 1}_0)$ & $(+, -)$ \\
    (${\bf 1}$, ${\bf 15}_0)$ & (+, +) \\
     (${\bf 1}$, ${\bf 6}_2)$ & $(+, -)$ \\
     (${\bf 1}$, ${\bf 6}_{-2})$ & $(+, -)$ \\ 
     (${\bf 1}$, ${\bf 1}_0)$ & (+, +) \\ 
     (${\bf 11}$, ${\bf 4}_{-1})$ & $(-, +)$ \\  
     (${\bf 11}$, $\overline{{\bf 4}}_{1})$ & $(-, -)$ \\ 
     (${\bf 1}$, ${\bf 4}_{-1})$ & $(-, +)$ \\ 
     (${\bf 1}$, $\overline{{\bf 4}}_{1})$ & $(-, -)$ \\
    \end{tabular}
     \caption{$Z_2$ parity assignments $(P_0, P_1)$ for various components in 6D $SO(20)$ gauge field $A_M$. }
       \label{Z25Dgauge}
       \end{table}
The $Z_2$ parity assignments for the various components in 6D $SO(20)$ gauge field are summarized in Table \ref{Z25Dgauge}. 
We can see from this table that the gauge fields of $SO(11) \times SU(4) \times U(1)$ only have $Z_2$ parity $(+, +)$, 
which means that they have massless modes. 
Furthermore, we have two massless mode from the sixth component of the gauge field $A_6$, 
$({\bf 11}, \overline{{\bf 4}}), ({\bf 1}, \overline{{\bf 4}})$ 
since the $Z_2$ parity is totally opposite to those of the 5D gauge field (\ref{Z26Dgauge}).  
However, these fields are non-singlet in a confining subgroup $Sp(4)$ in $SU(4)$. 
After all, no massless scalar field comes from $A_6$. 
Note that 5D gauge field $({\bf 1}, {\bf 15}_0)$ also confines and decoupled 
since no $Sp(4)$ singlet is included in ${\bf 15}$ of $SU(4)$  
as can be seen from the decomposition of ${\bf 15}$ of $SU(4)$ into $Sp(4)$ irreducible representations 
${\bf 15} = {\bf 10} + {\bf 5}$ \cite{Group1, Group2}. 
To summarize, the massless fields after compactification in the sixth dimension are the 5D $SO(11) \times U(1)$ gauge fields 
and three spinors in ${\bf 32}$ representation corresponding to three generations of the SM fermions. 

In order to obtain 4D theory by the orbifold compactification $S^1/Z_2$ in the fifth extra space, 
the following $Z_2$ parties are considered. 
\begin{align}
A_\mu(x_\mu, x_i-x_5) &= P'_i A_\mu(x_\mu, x_i + x_5) P_i^{'\dag}, \\
A_5(x_\mu, x_i-x_5) &= - P'_i A_6(x_\mu, x_i + x_5) P_i^{'\dag},
\label{Z25Dagauge}
\end{align}
where $Z_2$ matrices $P'_i~(i=0,1)$ at the fixed point $x_5=0, \pi R_5$ employed in \cite{HY} 
\begin{align}
P'_0 = I_{10, 1}, \quad P'_1 = I_{4, 7}. 
\end{align}
In terms of these $Z_2$ parities, the gauge symmetry breaking is given 
by\footnote{Here, $SU(4)$ including a confining subgroup $Sp(4)$ is neglected.}
\begin{align} 
SO(11) 
\times U(1) 
&\to SO(6) \times SO(4) \times U(1).  
\end{align}
As for the symmetry breaking to the SM gauge group, see \cite{HY}. 

A 5D $SO(11)$ massless gauge field is decomposed into $SO(6) \times SO(4)$ as follows.
\footnote{Here, $U(1)$ charges are neglected.} 
\begin{align}
{\bf 55}
=({\bf 15}, {\bf 1}) +({\bf 6}, {\bf 1}) + ({\bf 6}, {\bf 4}) + ({\bf 1}, {\bf 4}) + ({\bf 1}, {\bf 6}). 
\end{align}

\begin{table}[h]
    \centering
    \begin{tabular}{c|c}
   $(SO(6), SO(4))$& $(P'_0, P'_1)$ \\
    \hline
    (${\bf 15}$, ${\bf 1})$ & (+, +) \\
    (${\bf 6}$, ${\bf 1})$ & $(-, +)$ \\
     (${\bf 6}$, ${\bf 4})$ & $(+, -)$ \\ 
     (${\bf 1}$, ${\bf 4})$ & $(-, -)$ \\
     (${\bf 1}$, ${\bf 6})$ & (+, +) \\
             \end{tabular}
     \caption{$Z_2$ parity assignments $(P'_0, P'_1)$ for the various components 
     in 5D $SO(11)$ gauge field under $SO(6) \times SO(4)$. }
     \label{4DgaugeZ2}
\end{table}
The $Z_2$ parity assignments for the components in 5D $SO(11)$ gauge field is summarized in Table \ref{4DgaugeZ2}. 
We can see from the table that the gauge fields of $SO(6) \times SO(4)$ having $Z_2$ parity $(+, +)$ are massless.  
On the other hand, $({\bf 1}, {\bf 4})$ 
from $A_5$ are found to be massless and identified with the SM Higgs field. 
As for the fermion ${\bf 32}$, we can obtain three massless ${\bf 16}$ by following \cite{HY}. 

Yukawa coupling can be obtained from the gauge coupling 
through the fifth component of the gauge field $A_5$ in $SO(20)$ theory as follows. 
\begin{align}
{\bf 512} \cdot {\bf 190}_H \cdot {\bf 512} 
&\supset 
{\bf 32} \cdot {\bf 55}_H \cdot {\bf 32} + \cdots ({\rm 5D}~SO(11))\notag \\
&\supset {\bf 16} \cdot {\bf 10}_H \cdot {\bf 16} + \cdots, ({\rm 4D}~SO(10))
\end{align}
where the scalar field with subscript $H$ means 
that it includes the SM Higgs field\footnote{Here the $SO(10)$ expression does not mean 
that the 4D unbroken gauge symmetry is $SO(10)$. 
This is just an illustration of Yukawa coupling in terms of $SO(10)$ language.}. 
Also, the symbol $\supset$ implies that it extracts zero mode terms after compactification. 
In the first (second) line, the gauge coupling of 6D $SO(20)$ theory is decomposed into 5D $SO(11)$(4D $SO(10)$) symmetry. 
Since the gauge coupling is universal in flavor space, 
the fermion mass hierarchy, flavor mixing and CP violation cannot be explained as it stands. 
Improving these points is very challenging problem and requires nontrivial extensions. 

In summary, we have proposed a simple model of family unification based on the grand gauge-Higgs unification. 
The model is a 6D $SO(20)$ gauge theory with a single fermion in a spinorial representation. 
Remarkably, three generations of the SM fermions can be derived from a single fermion 
by assuming a subgroup $Sp(4)$ to be a confining gauge group. 
The 5D model obtained after compactification of the sixth dimension is 
a model of $SO(11) \times U(1)$ grand gauge-Higgs unification 
where the SM Higgs field is unified as the fifth component of the 5D gauge field
as well as only three generations of the SM fermions are massless in a fermion sector. 
Therefore, we believe that our model is the simplest model of family unification 
which further unifies the gauge fields and the Higgs field.  

There are issues to be addressed to discuss phenomenology based on this model. 
In family unified GHU models, Yukawa coupling constants are universal in flavor space 
since they are originated from the gauge coupling constant. 
How the fermion mass hierarchy, flavor mixing and CP violation are realized in our model is a common and important problem 
to be solved in models of family unification.  
One of the possible ways is to introduce the brane localized Yukawa interactions, 
which can provide the source of nontrivial flavor structure 
since the localized interactions does not need to respect the original symmetry.
It is also useful to introduce the $Z_2$ odd bulk masses, which makes the mode functions with different chiralities different 
and provides 4D Yukawa couplings with the suppression factor by the overlap integral of the mode functions.    
It will be interesting to investigate the fermion flavor structure in such bulk and brane Yukawa interaction systems. 

The issues of gauge coupling unification and proton decay analysis are also important and interesting. 
In particular, identifying the main decay mode in our model is crucial. 
In the model where the 6D bulk fields respecting the symmetry in the 6D bulk are introduced, 
the differences of the 1-loop gauge coupling running are expected to be almost logarithmic shown in \cite{MTY}, 
the unification might be realized at around $10^{14}$ GeV or so. 
Off course, we need to study it in detail since our model is two step breaking. 
In our model, the proton decay will occur via dimension six operators of $X, Y$ gauge boson and $A_{5,6}$ scalar field exchange. 
If the unification scale is around $10^{14}$ GeV as mentioned above, 
the suppression of the proton decay is not enough. 
However, we guess that some suppression factors come from the overlap integral of mode functions in deriving dimension six operators. 
If the wave functions of quarks and leptons are separately localized in extra spaces, 
the proton decay can be easily suppressed by the overlap integral of the wave functions \cite{AS}. 
Therefore, we expect that the proton decay constraints can be satisfied. 
In a case that the effects of the localization are not enough, 
it might be possible to suppress them by some symmetries (such as the discrete symmetry discussed in \cite{UED}, for instance).

Although two step compactifications of 6D and the confining gauge dynamics of $Sp(4)$ were assumed in this paper, 
a model building based on a compactification on an orbifold $T^2/Z_N$ 
and that without any confining gauge group would be more interesting. 
These are left for our future work. 

\subsection*{Acknowledgments}

This work was supported in part by JSPS KAKENHI Grant-in-Aid for Scientific Research (C) No.~25K07304(NM) 
and Grant-in-Aid for JSPS Research Fellows (RN).  


\end{document}